\documentclass[prd,showpacs,showkeys,nofootinbib,twocolumn]{revtex4-1}
\usepackage{amsmath,amsfonts,amssymb}
\usepackage[colorlinks]{hyperref}
\usepackage{bm}
\usepackage{epsf}
\usepackage{color}
\usepackage{dcolumn}
\usepackage{bm}
\everymath{\displaystyle}

\begin{document} 

\title{Gravitational waves in Poincar\'e gauge gravity theory}

\author{\firstname{Yuri N.}~\surname{Obukhov}}
\email{obukhov@ibrae.ac.ru}
\affiliation{Theoretical Physics Laboratory, Nuclear Safety Institute,
Russian Academy of Sciences, B. Tulskaya 52, 115191 Moscow, Russia}


\begin {abstract}
We derive the exact gravitational wave solutions in a general class of quadratic Poincar\'e gauge gravity models. The Lagrangian includes all possible linear and quadratic invariants constructed from the torsion and the curvature, including the parity odd terms. The ansatz for the gravitational wave configuration and the properties of the wave solutions are patterned following the corresponding ansatz and the properties of the plane-fronted electromagnetic wave.  
\end{abstract}

\pacs{04.50.-h, 04.20.Jb, 04.30.-w} \maketitle

\section{Introduction}

Einstein's general relativity (GR) theory used the beautiful and powerful methods of differential geometry to describe the gravitational phenomena in terms of the geometrical properties of the four-dimensional spacetime manifold. Technically, the metric structure and the Riemannian curvature formed the core of GR as a theory of a macroscopic gravitational field. Later, it was recognized that the three other physical interactions (electromagnetic, weak and strong) can also be geometrized by making use of the Yang-Mills gauge-theoretic approach, and a natural question arose whether the gravity theory can be consistently formulated in the the gauge framework. The answer is not that simple as it might appear at the first sight. The subtle point is that the Standard Model is based on the fundamental symmetry groups acting in the internal spaces, whereas the gravity is obviously related with the symmetry of the spacetime itself. In Einstein's GR, the central role is played by the group of the local spacetime translations (diffeomorphisms), which is directly reflected in the fact the corresponding translational Noether current --the energy-momentum tensor-- is the physical source of the gravitational field, see \cite{Feyn}. At the same time, the Poincar\'e group (a semidirect product of the Lorentz group times the group of translations) plays a significant role in the theoretical high energy physics: recall that the elementary particles are classified by mass and spin in the representation theory of the Poincar\'e group.

 Not surprisingly, the Poincar\'e group underlies the development of the gauge theory of gravity \cite{rmp,MAG,Blag,reader}. There is a general belief that the gauge gravity model may correctly describe the gravitational phenomena at microscopic scales (and, likewise, at an early stage of universe's evolution), whereas GR arises as a certain macroscopic limit of the Poincar\'e gauge gravity. Indeed, it was shown very early \cite{Sciama,Kibble,HonnefH,Tra} that a viable extension of GR based on the Poincar\'e group can be constructed, in which the energy-momentum and spin currents arise as the sources of the gravitational field. The corresponding Riemann-Cartan spacetime geometry is characterized by a nontrivial torsion which is coupled to spin current, along with the metric coupled to the energy-momentum current. This is perfectly consistent with the semidirect structure of the Poincar\'e group, and the Noether-Lagrange formalism identifies the energy-momentum tensor (``translational'' Noether current, corresponding to the group of translations) and the tensor of spin (``rotational'' current, corresponding to the Lorentz group) with the two physical matter sources of the Poincar\'e gauge field. The principle of the local invariance relates the existence of the gauge fields to the corresponding Noether currents. 

The early history of the theory of gravity with torsion includes the development by \'Elie Cartan of the first gravitational model \cite{Cartan} which then inspired Einstein so much in his search for a unified field theory (see the historic account in \cite{Goenner}). Weyl, Einstein, Eddington, Schr\"odinger, Cosserats, and Kroener \emph{et al} \cite{essay} made significant progress in a further understanding of the non-Riemannian geometries and contributed to the development of the physical theories based on such generalized geometries. The relation of the torsion (more generally, of the Riemann-Cartan geometry of a spacetime manifold with curvature and torsion) to the gravitational physics is now well understood in the framework of the Poincar\'e gauge theory \cite{Sciama,Kibble,HonnefH,MAG,Blag,reader}. The Einstein-Cartan gravitational theory \cite{Tra,rmp}, as the closest viable extension of GR, agrees with physical observations on the macroscopic scales, and in particular it is consistent with all classical gravity experimental tests within the Solar System. The contact spin-torsion interaction is responsible for the deviation from GR on the microscopic scales in the high energy particle experiments with polarized matter and for the modification of early stages of cosmological evolution at extremely high densities of matter. 

In this paper we discuss the plane gravitational waves with nontrivial torsion and curvature for the class of the Poincar\'e gauge gravity models with the general Yang-Mills type quadratic Lagrangian. Waves belong to the most fundamental physical phenomena, and a systematic study of the corresponding solutions potentially contributes to the development of a field-theoretic model under consideration. Therefore, the construction and comparative analysis of the wave solutions in different gravitational models is an important task which can clarify the physical contents of the model and the relations between the microscopic and macroscopic gravitational theories (in particular, between the general relativity, Poincar\'e gauge gravity and the metric-affine gravity).  

The plan of the paper is as follows. In Sec.~\ref{PG} we present a condensed introduction to the Poincar\'e gauge gravity theory. Only the basic definitions are given and the main structures are described, whereas the details of the gauge approach in the gravitational theory can be found in the relevant literature \cite{rmp,MAG,Blag,reader}. In Sec.~\ref{GW} we recall the properties of a plane-fronted electromagnetic wave and use this to formulate the appropriate ansatz for the gravitational wave in the Poincar\'e gauge gravity. The properties of the resulting curvature and torsion 2-forms are investigated. In Sec.~\ref{FE} the set of differential equations for the wave variables is derived. Although the original field equations of the Poincar\'e gauge theory are highly nonlinear, the variables that describe wave's profile satisfy a system of linear equations of Helmholtz and/or screened Laplace equations. Solutions of this system are investigated. Finally, the conclusions are outlined in Sec.~\ref{DC}.

Our basic notation and conventions are consistent with \cite{MAG}. In particular, Greek indices $\alpha, \beta, \dots = 0, \dots, 3$, denote the anholonomic components (for example, of a coframe $\vartheta^\alpha$), while the Latin indices $i,j,\dots =0,\dots, 3$, label the holonomic components ($dx^i$, e.g.). The anholonomic vector frame basis $e_\alpha$ is dual to the coframe basis in the sense that $e_\alpha\rfloor\vartheta^\beta = \delta_\alpha^\beta$, where $\rfloor$ denotes the interior product. The volume 4-form is denoted $\eta$, and the $\eta$-basis in the space of exterior forms is constructed with the help of the interior products as $\eta_{\alpha_1 \dots\alpha_p}:= e_{\alpha_p}\rfloor\dots e_{\alpha_1}\rfloor\eta$, $p=1,\dots,4$. They are related to the $\theta$-basis via the Hodge dual operator $^*$, for example, $\eta_{\alpha\beta} = {}^*\!\left(\vartheta_\alpha\wedge\vartheta_\beta\right)$. The Minkowski metric $g_{\alpha\beta} = {\rm diag}(+1,-1,-1,-1)$.

\section{Poincar\'e gauge gravity: crash course}\label{PG}

The gravitational field is described by the coframe $\vartheta^\alpha = e^\alpha_i dx^a$ and connection $\Gamma_\alpha{}^\beta = \Gamma_{i\alpha}{}^\beta dx^i$ 1-forms. The translational and rotational field strengths read 
\begin{eqnarray}
T^\alpha &=& D\vartheta^\alpha = d\vartheta^\alpha +\Gamma_\beta{}^\alpha\wedge
\vartheta^\beta,\label{Tor}\\ \label{Cur}
R_\alpha{}^\beta &=& d\Gamma_\alpha{}^\beta + \Gamma_\gamma{}^\beta\wedge\Gamma_\alpha{}^\gamma.
\end{eqnarray}
As usual, the covariant differential is denoted $D$.

The gravitational Lagrangian 4-form 
\begin{equation}
V = V(\vartheta^{\alpha}, T^{\alpha}, R_{\alpha}{}^{\beta})\label{lagrV}
\end{equation}
is (in general) an arbitrary function of the geometrical variables. Its variation reads
\begin{eqnarray}
\delta V &=& \delta\vartheta^{\alpha}\wedge{\cal E}_\alpha
+ \delta\Gamma_\alpha{}^{\beta}\wedge{\cal C}^\alpha{}_\beta \nonumber\\
&& - \,d\left[\delta\vartheta^{\alpha}\wedge H_\alpha + \delta\Gamma_{\alpha}{}^{\beta}
\wedge H^{\alpha}{}_{\beta}\right].\label{deltaV}
\end{eqnarray}
Here, as usual, we introduce the Poincar\'e {\it gauge field momenta} 2-forms
\begin{equation} 
H_{\alpha} := -\,{\frac{\partial V}{\partial T^{\alpha}}}\,,\qquad  
H^{\alpha}{}_{\beta} := -\,{\frac{\partial V}{\partial R_{\alpha}{}^{\beta}}}\,,\label{HH}
\end{equation}  
define the $3$--forms of the {\it canonical} energy--momentum and 
spin for the gravitational gauge fields:
\begin{equation} 
E_{\alpha} := {\frac{\partial V}{\partial\vartheta^{\alpha}}},\qquad
E^\alpha{}_\beta := {\frac{\partial V}{\partial\Gamma_\alpha{}^\beta}} = 
- \vartheta^{[\alpha}\wedge H_{\beta]}\,, \label{EE}
\end{equation}
and ultimately find the variational derivatives with respect to the gravitational
field potentials
\begin{eqnarray}
{\mathcal E}_\alpha &:=& {\frac{\delta V}{\delta\vartheta^{\alpha}}} = 
- DH_{\alpha}  +E_{\alpha}, \label{dVt}\\ 
{\mathcal C}^\alpha{}_\beta &:=& {\frac{\delta V}{\delta\Gamma_\alpha{}^\beta}} 
= - DH^\alpha{}_\beta + E^\alpha{}_\beta\,.\label{dVG}
\end{eqnarray}
Diffeomorphism invariance yields the Noether identities
\begin{eqnarray}
E_{\alpha} &\equiv& e_{\alpha}\rfloor V + (e_{\alpha}\rfloor T^{\beta})\wedge H_{\beta} 
+ (e_{\alpha}\rfloor R_{\beta}{}^{\gamma})\wedge H^{\beta}{}_{\gamma},\label{Ea}\\
D\,{\mathcal E}_\alpha &\equiv& (e_{\alpha}\rfloor T^{\beta})\wedge{\mathcal E}_\beta 
+ (e_{\alpha}\rfloor R_{\beta}{}^{\gamma})\wedge\,{\mathcal C}^\beta{}_\gamma,\label{1st}
\end{eqnarray}
whereas the local Lorentz invariance results in the Noether identity
\begin{equation}
2D{\mathcal C}^\alpha{}_\beta + \vartheta^\alpha\wedge{\mathcal E}_\beta 
- \vartheta_\beta\wedge{\mathcal E}^\alpha \equiv 0\,.\label{2nd}
\end{equation}

\subsection{Gravitational field equations: general case}

Quite generally, we describe matter by the set of $p$-forms $\psi^A$ (we do not need to specify the value of $p$ and the range of the indices $A, B, \dots$ which label components of matter variables). 

The field equations for the system of interacting matter $\psi^A$ and gravitational fields
$\vartheta^\alpha$ and $\Gamma_\alpha{}^\beta$ are derived from the total Lagrangian
\begin{equation}
V(\vartheta^{\alpha}, T^{\alpha}, R_{\alpha}{}^{\beta}) + {\frac 1c}
L(\psi^A, D\psi^A, \vartheta^{\alpha}, T^{\alpha}, R_{\alpha}{}^{\beta}).\label{Ltot}
\end{equation}
Independent variation with respect to $\psi^A$, $\vartheta^\alpha$
and $\Gamma_\alpha{}^\beta$ yields
\begin{eqnarray}
{\frac{\delta L}{\delta\psi^A}} &=& {\frac {\partial  L}{\partial\psi^A}} - (-1)^{p}D\,
{\frac {\partial L}{\partial (D\psi^A)}} = 0\,,\label{Pmat}\\ 
{\mathcal E}_\alpha &=& {\frac 1c}{\mathfrak T}_{\alpha}\,,\label{Peq1}\\
{\mathcal C}^\alpha{}_\beta &=& {\frac 12}{\mathfrak S}^{\alpha}{}_\beta\,.\label{Peq2}
\end{eqnarray} 
The {\it matter currents} in (\ref{Peq1}) and (\ref{Peq2}) are the canonical 
energy-momentum 3-form and the canonical spin density 3-form of matter. They are given by
\begin{eqnarray}
{\mathfrak T}_{\alpha} &:=& -\,{\frac {\delta L}{\delta\vartheta^{\alpha}}} =  
- \,{\frac {\partial L}{\partial\vartheta^{\alpha}}} 
- D\,{\frac {\partial L}{\partial T^{\alpha}}}\, ,\label{sigC0}\\
c{\mathfrak S}^\alpha{}_\beta &:=& -\,2{\frac {\delta L}{\delta\Gamma_\alpha{}^\beta}} =   
(\rho^A{}_B)^\alpha{}_\beta\psi^B\wedge{\frac {\partial L} {\partial (D\psi^A)}}\nonumber\\
&& -\,\vartheta^\alpha\wedge {\frac {\partial L}{\partial T^\beta}} +
\vartheta_\beta\wedge {\frac {\partial L}{\partial T_\alpha}} 
-2D{\frac {\partial L}{\partial R_\alpha{}^\beta}}\,.\label{spin0}
\end{eqnarray}

\subsection{Einstein-Cartan model}

The Einstein-Cartan theory is based on the Hilbert-Einstein Lagrangian
\begin{equation}
V_{HE} = {\frac {1}{2\kappa c}}\eta_{\alpha\beta}\wedge R^{\alpha\beta}.\label{LHE}
\end{equation}
Here $\kappa = {\frac {8\pi G}{c^4}}$ is Einstein's gravitational constant.

For the Lagrangian (\ref{LHE}) we find from (\ref{HH}), (\ref{EE}) and (\ref{Ea}):
\begin{eqnarray}\label{HHEC}
H_\alpha = 0,\quad H^\alpha{}_\beta = -\,{\frac {1}{2\kappa c}}\eta^\alpha{}_\beta,\\
E_\alpha = {\frac {1}{2\kappa c}}\eta_{\alpha\beta\gamma}\wedge R^{\beta\gamma},
\quad E^\alpha{}_\beta = 0.\label{EEEC}
\end{eqnarray}
As a result, ${\cal E}_\alpha = {\frac {1}{2\kappa c}}\eta_{\alpha\beta\gamma}\wedge R^{\beta\gamma}$ and ${\cal C}^\alpha{}_\beta = {\frac {1}{2\kappa c}}\eta^\alpha{}_{\beta\gamma}\wedge T^{\gamma}$, and hence the Einstein-Cartan field equations read
\begin{equation}
 {\frac {1}{2}}\eta_{\alpha\beta\gamma}\wedge R^{\beta\gamma} = \kappa\,{\mathfrak T}_\alpha,\qquad
\eta^\alpha{}_{\beta\gamma}\wedge T^{\gamma} = \kappa c\,{\mathfrak S}^\alpha{}_\beta.\label{ECeq}
\end{equation}

\subsection{Quadratic Poincar\'e gravity models}

The torsion 2-form can be decomposed into the 3 irreducible parts, whereas the curvature 2-form has 6 irreducible pieces. Their definition is presented in Appendix~\ref{irreducible}.

The general quadratic model is described by the Lagrangian 4-form that contains all possible quadratic 
invariants of the torsion and the curvature:
\begin{eqnarray}
V &=& {\frac {1}{2\kappa c}}\Big\{\Big(a_0\eta_{\alpha\beta} + {\frac 1\xi}
\vartheta_\alpha\wedge\vartheta_\beta\Big)\wedge R^{\alpha\beta} \nonumber\\
&& -\,T^\alpha\wedge\sum_{I=1}^3
\left[a_I\,{}^*({}^{(I)}T_\alpha) + \overline{a}_I\,{}^{(I)}T_\alpha\right]\Big\}\nonumber\\
&& - \,{\frac 1{2\rho}}R^{\alpha\beta}\wedge\sum_{I=1}^6 \left[b_I\,{}^*({}^{(I)}\!R_{\alpha\beta}) 
+ \overline{b}_I\,{}^{(I)}\!R_{\alpha\beta}\right].\label{LRT}
\end{eqnarray}
The Lagrangian has a clear structure: the first line is {\it linear} in the curvature, the second line collects {\it torsion quadratic} terms, whereas the third line contains the {\it curvature quadratic} invariants. Furthermore, each line is composed of the parity even pieces (first terms on each line), and the parity odd parts (last terms on each line). The dimensionless constant $\xi$ is called a Barbero-Immirzi parameter, and the linear part of the Lagrangian -- the first line in (\ref{LRT}) -- describes what is known in the literature as the Einstein-Cartan-Holst model. 

There was considerable interest in the gravity models with parity odd Lagrangians, especially in cosmology, during the recent time: see \cite{Obukhov:1989,Chen,Ho1,Ho2,Ho3,Diakonov,Baekler1,Baekler2,Karananas}, e.g.

The Lagrangian contains sets of coupling constants: $\rho$, $a_1, a_2, a_3$ and $\overline{a}_1, \overline{a}_2, \overline{a}_3$, $b_1, \cdots, b_6$ and $\overline{b}_1, \cdots, \overline{b}_6$. The overbar denotes the constants responsible for the parity odd interaction. We have the dimension $[{\frac 1\rho}] = [\hbar]$, whereas $a_I$, $\overline{a}_I$, $b_I$ and $\overline{b}_I$ are dimensionless. Moreover, not all of these constants are independent: we take $\overline{a}_2 = \overline{a}_3$, $\overline{b}_2 = \overline{b}_4$ and $\overline{b}_3 = \overline{b}_6$ because some of terms in (\ref{LRT}) are the same: 
\begin{eqnarray}\label{T23}
T^\alpha\wedge{}^{(2)}T_\alpha = T^\alpha\wedge{}^{(3)}T_\alpha = {}^{(2)}T^\alpha\wedge{}^{(3)}T_\alpha,\\
R^{\alpha\beta}\wedge{}^{(2)}\!R_{\alpha\beta} = R^{\alpha\beta}\wedge{}^{(4)}\!R_{\alpha\beta} 
= {}^{(2)}\!R^{\alpha\beta}\wedge{}^{(4)}\!R_{\alpha\beta},\label{R24} \\ 
R^{\alpha\beta}\wedge{}^{(3)}\!R_{\alpha\beta} = R^{\alpha\beta}\wedge{}^{(6)}\!R_{\alpha\beta} 
= {}^{(3)}\!R^{\alpha\beta}\wedge{}^{(6)}R_{\alpha\beta}.\label{R36}
\end{eqnarray}
whereas $T^\alpha\wedge{}^{(1)}T_\alpha = {}^{(1)}T^\alpha\wedge{}^{(1)}T_\alpha$ and 
$R^{\alpha\beta}\wedge{}^{(1)}\!R_{\alpha\beta} = {}^{(1)}\!R^{\alpha\beta}\wedge{}^{(1)}
\!R_{\alpha\beta}$ and $R^{\alpha\beta}\wedge{}^{(5)}\!R_{\alpha\beta} = {}^{(5)}\!R^{\alpha\beta}
\wedge{}^{(5)}\!R_{\alpha\beta}$. One can prove these relations directly from the 
definitions (\ref{iT2})-(\ref{iT1}) and (\ref{curv2})-(\ref{curv1}).

For completeness, we included the dimensionless constant $a_0$. This allows for the special case $a_0 = 0$ of the purely quadratic model without the Hilbert-Einstein linear term in the Lagrangian. In the Einstein-Cartan model (\ref{LHE}), one puts $a_0 = 1$.

For the Lagrangian (\ref{LRT}) from (\ref{HH}), (\ref{EE}) and (\ref{Ea}) 
we derive the gauge gravitational field momenta 
\begin{eqnarray}
H_\alpha &=& {\frac 1{\kappa c}}\sum_{I=1}^3\left[a_I\,{}^*({}^{(I)}T_\alpha) 
+ \overline{a}_I\,{}^{(I)}T_\alpha\right], \label{HaRT}\\
H^\alpha{}_\beta &=& -\,{\frac {1}{2\kappa c}}\left(a_0\,\eta^\alpha{}_\beta + {\frac 1\xi}
\vartheta^\alpha\wedge\vartheta_\beta\right)\nonumber\\
&& +\,{\frac 1{\rho}}\sum_{I=1}^6\left[b_I\,{}^*({}^{(I)}\!R^\alpha{}_\beta) 
+ \overline{b}_I\,{}^{(I)}\!R^\alpha{}_\beta\right],\label{HabRT}
\end{eqnarray}
and the canonical energy-momentum and spin currents of the gravitational field
\begin{eqnarray}
E_\alpha &=& {\frac {1}{2\kappa c}}\left(a_0\,\eta_{\alpha\beta\gamma}\wedge R^{\beta\gamma} + 
{\frac 2\xi}R_\alpha{}^\beta\wedge\vartheta_\beta\right)\nonumber\\
&& + \,{\frac 12}\left[(e_\alpha\rfloor T^\beta)\wedge H_\beta - T^\beta\wedge 
e_\alpha\rfloor H_\beta\right]\nonumber\\
&& + \,{\frac 1{2\rho}}\left[(e_\alpha\rfloor R_\beta{}^\gamma)\wedge h^\beta{}_\gamma 
- R_\beta{}^\gamma\wedge e_\alpha\rfloor h^\beta{}_\gamma\right],\label{EaRT}\\
E^\alpha{}_\beta &=& {\frac 12}\left(H^\alpha\wedge\vartheta_\beta 
- H_\beta\wedge\vartheta^\alpha\right).\label{EabRT}
\end{eqnarray}
For convenience, we introduce the following 2-forms which are linear functions 
of the torsion and the the curvature, respectively, by
\begin{eqnarray}
h_\alpha &=& \kappa cH_\alpha = \sum_{I=1}^3\left[a_I\,{}^*({}^{(I)}T_\alpha) 
+ \overline{a}_I\,{}^{(I)}T_\alpha\right],\label{HlaQT}\\
h^\alpha{}_\beta &=& \sum_{I=1}^6\left[b_I\,{}^*({}^{(I)}\!R^\alpha{}_\beta) 
+ \overline{b}_I\,{}^{(I)}\!R^\alpha{}_\beta\right].\label{hR}
\end{eqnarray}
By construction, the former object has the dimension of a length, $[h_\alpha] = [\ell]$, whereas the latter 2-form is obviously dimensionless, $[h^\alpha{}_\beta] = 1$. It is straightforward to evaluate ${\cal E}_\alpha = -\,DH_\alpha + E_\alpha$ and ${\cal C}^\alpha{}_\beta = -\,DH^\alpha{}_\beta + E^\alpha{}_\beta$. The resulting Poincar\'e gravity field equations (\ref{Peq1}) and (\ref{Peq2}) then read 
\begin{eqnarray}
{\frac {a_0}2}\eta_{\alpha\beta\gamma}\wedge R^{\beta\gamma} + {\frac 1\xi}R_\alpha{}^\beta
\wedge\vartheta_\beta  && \nonumber\\ 
+ \,q^{(T)}_\alpha + \ell_\rho^2\,q^{(R)}_\alpha - Dh_\alpha &=& \kappa\,{\mathfrak T}_\alpha,\label{ERT1}\\
a_0\,\eta^\alpha{}_{\beta\gamma}\wedge T^{\gamma} + {\frac 1\xi}\left(T^\alpha\wedge\vartheta_\beta
- T_\beta\wedge\vartheta^\alpha \right) && \nonumber\\ 
+ \,h^\alpha\wedge\vartheta_\beta - h_\beta\wedge\vartheta^\alpha 
- 2\ell_\rho^2\,Dh^\alpha{}_\beta &=& \kappa c\,{\mathfrak S}^\alpha{}_\beta.\label{ERT2}
\end{eqnarray}
We introduced here the following 3-forms which
are quadratic in the torsion and in the curvature, respectively:
\begin{eqnarray}
q^{(T)}_\alpha &=& {\frac 12}\left[(e_\alpha\rfloor T^\beta)\wedge h_\beta - T^\beta\wedge 
e_\alpha\rfloor h_\beta\right],\label{qa}\\
q^{(R)}_\alpha &=& {\frac 12}\left[(e_\alpha\rfloor R_\beta{}^\gamma)\wedge h^\beta{}_\gamma 
- R_\beta{}^\gamma\wedge e_\alpha\rfloor h^\beta{}_\gamma\right].\label{qaR}
\end{eqnarray}
The former object carries the dimension of length $[q^{(T)}_\alpha] = [\ell]$, and the latter one has the dimension of the inverse length, $[q^{(R)}_\alpha] = [1/\ell]$.

The contribution of the curvature square terms in the Lagrangian (\ref{LRT}) to the gravitational field dynamics in the equations (\ref{ERT1}) and (\ref{ERT2}) is characterized by the parameter
\begin{equation}
\ell_\rho^2 = {\frac {\kappa c}{\rho}}.\label{lr}
\end{equation}
Since $[{\frac 1\rho}] = [\hbar]$, this new coupling parameter has the dimension of the area, $[\ell_\rho^2] = [\ell^2]$.

\section{Gravitational waves in Poincar\'e gauge gravity}\label{GW}

Gravitational waves are of fundamental importance in physics, and recently the purely theoretical research in this area was finally supported by the first experimental evidence \cite{Abbott1,Abbott2}. A general overview of the history of this fascinating subject can be found in \cite{flan,schutz,CNN}. 

The plane-fronted gravitational waves represent an important class of exact solutions which generalize the basic properties of electromagnetic waves in flat spacetime to the case of curved spacetime geometry. The extensive investigations of these solutions in Einstein's general relativity have a long and rich history, see, e.g., \cite{Brink1,Brink2,Brink3,rosen1937,einrosen,rosen1956,rosen1958,Virb1,Virb2,bondi0,bondi1,peres,pen1,pen2,Kom1,Kom2,Jordan1,Jordan2,kundt,curr,schim,AT,piran,MashQ,torre,cropp1,cropp2,coley12,mcnutt,Barnett,griff,vdz,exact}. Various generalizations of plane-fronted wave solutions were reported for the Poincar\'e gauge gravity \cite{adam,chen,sippel,vadim,singh,babu}, for the teleparallel gravity \cite{tele}, for the generalized Einstein theories \cite{gurses,lovelock,Mohseni}, for supergravity \cite{sg1,sg2,sg3,sg4,sg5}, as well as, more recently, for the superstring theories \cite{gimon,ark1,ark2,str1,str2,str3,str4}. The higher-dimensional generalizations of the gravitational wave solutions were discussed, in particular, in \cite{sokol,coley1,coley2,hervik,ndim}. Moreover, it was discovered \cite{ppmag,dirk1,dirk2,king,vas1,vas2,vas3,pasic1,pasic2} that gravitational plane wave solutions are also admitted in the metric-affine theory of gravity (MAG) with the propagating torsion and nonmetricity fields.

\subsection{Electromagnetic plane wave}

The key for the description of a plane wave on a spacetime manifold is the null 
shear-free geodetic vector field $k^i$:
\begin{equation}
k_ik^i = 0,\qquad k^j\nabla_jk^i = 0.\label{kk0}
\end{equation}
The plane electromagnetic wave is given by the field strength tensor $F_{ij}$ that solves the Maxwell equations and satisfies
\begin{equation}
k^jF_{ij} = 0,\qquad k_{[i}F_{jk]} = 0,\qquad F_{ij}F^{ij} = 0.\label{kF0}
\end{equation} 

In exterior language, the wave 1-form $k = d\varphi$ arises from the phase function $\varphi$, and the wave covector is $k_\alpha = e_\alpha\rfloor k$. The properties (\ref{kk0}) and (\ref{kF0}) are then recast into
\begin{eqnarray}
k\wedge{}^\ast\!k = 0,\qquad k\wedge{}^\ast\!Dk^\alpha = 0,\label{kk1}\\
k\wedge{}^\ast\!F = 0,\qquad k\wedge F = 0,\qquad F\wedge{}^\ast\!F = 0.\label{kF1}
\end{eqnarray}

The actual form of the wave configurations depends on the Lagrangian of the electromagnetic field. For example, in Maxwell's theory in the flat Minkowski spacetime (specializing to the case $e^\alpha_i = \delta^\alpha_i, \Gamma_i{}^{\alpha\beta} = 0$) the electromagnetic plane wave is given by
\begin{equation}
F = k\wedge a,\qquad {\rm or}\qquad F_{ij} = k_ia_j - k_ja_i.\label{Fij}
\end{equation}
Here the wave covector $k_i$ is constant, and the polarization 1-form $a$ (covector $a_i$) depends only on the phase $\varphi = k_ix^i$: $a_i = a_i(\varphi)$. Polarization 1-form satisfies the orthogonality relation
\begin{equation}
k\wedge{}^\ast\!a = 0,\qquad {\rm or}\qquad k^ia_i = 0.\label{ai}
\end{equation}
One can check that Maxwell's equations $dF = 0$ and $d{}^\ast\!F = 0$ ($\partial_jF^{ij} = 0$, and $\partial_{[j}F_{kl]} = 0$) are satisfied, and the field invariants are trivial $F_{ij}F^{ij} = 0, \,\eta^{ijkl}F_{ij}F_{kl} = 0$.

In a similar way, we expect that a plane gravitational wave is described by the corresponding gravitational field strength, the curvature 2-form $R^{\alpha\beta}$, with the properties
\begin{equation}
k\wedge{}^\ast\!R^{\alpha\beta} = 0,\qquad k\wedge R^{\alpha\beta} = 0,\qquad 
R^{\alpha\beta}\wedge{}^\ast\!R^{\rho\sigma} = 0,\label{kRW0}
\end{equation}
generalizing (\ref{kF1}). In components, $R^{\alpha\beta} = {\frac 12}R_{ij}{}^{\alpha\beta}\,dx^i\wedge dx^j$:
\begin{equation}
k^jR_{ij}{}^{\alpha\beta} = 0,\qquad k_{[i}R_{jk]}{}^{\alpha\beta} = 0,\qquad 
R_{ij}{}^{\alpha\beta}R^{ij}{}_{\mu\nu} = 0.\label{kR0}
\end{equation}
This is in a close analogy to (\ref{kF0}).

\subsection{Plane wave ansatz in gauge gravity}

Let us now describe the plane wave ansatz in  Poincar\'e gauge gravity. As a first step, we divide the local coordinates into two groups: $x^i= (x^a, x^A)$, where $x^a = (x^0 = \sigma, x^1 = \rho)$ and $x^A = (x^2,x^3)$. Hereafter the indices from the beginning of the Latin alphabet $a,b,c... = 0,1$, whereas the capital Latin indices run $A,B,C... = 2,3$. 

The coframe 1-form is chosen as
\begin{eqnarray}
\vartheta^{\widehat 0} &=& {\frac 12}(U + 1)d\sigma + {\frac 12}\,d\rho,\label{cof0}\\ 
\vartheta^{\widehat 1} &=& {\frac 12}(U - 1)d\sigma + {\frac 12}\,d\rho,\label{cof1}\\
\vartheta^{\widehat A} &=& dx^A,\qquad A = 2,3.\label{cof23}
\end{eqnarray}
Here $U = U(\sigma, x^A)$. As a result, the line element reads
\begin{equation}
ds^2 = d\sigma d\rho + Ud\sigma^2 - \delta_{AB}dx^Adx^B.\label{ds_2}
\end{equation}
It is common to use the so-called null (or semi-null) Minkowski metric for the discussion of the gravitational waves. We, however, throughout this paper make use of the standard diagonal metric $g_{\alpha\beta} = {\rm diag}(+1,-1,-1,-1)$. 

Following the analogy with the electromagnetism, we now introduce a crucial object: the wave 1-form $k$. We define the latter as 
\begin{equation}
k := d\sigma = \vartheta^{\widehat 0} - \vartheta^{\widehat 1}.\label{kdef}
\end{equation}
By construction, we have $k\wedge{}^\ast\!k = 0$. As before, the wave covector is $k_\alpha = e_\alpha\rfloor k$. Its (anholonomic) components are thus $k_\alpha = (1, -1, 0, 0)$ and $k^\alpha = (1, 1, 0, 0)$. Hence, this is a null vector field, $k_\alpha k^\alpha = 0$. 

For the local Lorentz connection 1-form, we assume
\begin{equation}
\Gamma_\alpha{}^\beta = k\left(k_\alpha W^\beta - k^\beta W_\alpha\right),\label{conW}
\end{equation}
where the new vector variable $W^\alpha = W^\alpha(\sigma, x^A)$. In addition, we assume the orthogonality 
\begin{equation}
k_\alpha W^\alpha = 0.\label{kW0}
\end{equation}
This is guaranteed if we choose 
\begin{equation}\label{Wa0}
W^\alpha = \begin{cases}W^a = 0,\qquad\qquad\qquad a = 0,1, \\
W^A = W^A(\sigma, x^B),\qquad A = 2,3.\end{cases}
\end{equation}
Here $W^2(\sigma, x^B)$ and $W^3(\sigma, x^B)$ are the two unknown functions. 

In other words, the ansatz for the Poincar\'e gauge potentials -- coframe (\ref{cof0})-(\ref{cof23}) and connection (\ref{conW}) -- is described by the three variables $U = U(\sigma, x^B)$ and $W^A = W^A(\sigma, x^B)$. These functions determine wave's profile and their explicit form should be found from the gravitational field equations.

One immediately verifies that the wave 1-form is closed, and the wave covector is constant:
\begin{equation}
dk = 0,\qquad dk_\alpha = 0,\qquad Dk_\alpha = 0.\label{dk0}
\end{equation}
Taking this into account, we straightforwardly compute the torsion and the curvature 2-forms:
\begin{eqnarray}
T^\alpha &=& -\,k\wedge k^\alpha\,\Theta,\label{torW}\\
R_\alpha{}^\beta &=& -\,k\wedge\left(k_\alpha \Omega^\beta - k^\beta \Omega_\alpha\right),\label{curW}
\end{eqnarray}
where we introduced the 1-forms 
\begin{eqnarray}
\Theta &:=& {\frac 12}\,\underline{d}\,U + W_\alpha\vartheta^\alpha,\label{THW}\\
\Omega^\alpha &:=& \underline{d}\,W^\alpha. \label{OMW}
\end{eqnarray}
The differential $\underline{d}$ acts in the transversal 2-space spanned by $x^A = (x^2, x^3)$:
\begin{equation}
 \underline{d} := \vartheta^Ae_A\rfloor d = dx^A\partial_A,\qquad A = 2,3.\label{ud}
\end{equation}
Although the geometry of the transversal 2-space spanned by $x^A = (x^2, x^3)$ is fairly simple, it is convenient to describe it explicitly. It is a flat Euclidean space with the volume 2-form $\underline{\eta} = {\frac 12}\eta_{AB}\vartheta^A\wedge\vartheta^B = dx^2\wedge dx^3$, where $\eta_{AB} = -\,\eta_{BA}$ is the 2-dimensional Levi-Civita tensor (with $\eta_{23} = 1$). The volume 4-form of the spacetime manifold then reads $\eta = \vartheta^{\widehat 0}\wedge \vartheta^{\widehat 1}\wedge\vartheta^{\widehat 2}\wedge\vartheta^{\widehat 3} = {\frac 12}k\wedge d\rho\wedge\underline{\eta}$. For the wave 1-form we find the remarkable relation 
\begin{equation}
{}^*k = -\,k\wedge\underline{\eta}.\label{dualk}
\end{equation}
We will denote the geometrical objects on the transversal 2-space by underlining them; for example, a 1-form $\underline{\phi} = \phi_A\vartheta^A$. The Hodge duality on this space is defined as usual via ${}^{\underline{*}}\vartheta_A = \underline{\eta}_A = e_A\rfloor \underline{\eta} = \eta_{AB}\vartheta^B$. With the help of (\ref{dualk}), we can verify
\begin{equation}
{}^*(k\wedge\underline{\phi}) = k\wedge{}^{\underline{*}}\underline{\phi}.\label{kphi}
\end{equation}

The new objects (\ref{THW}) and (\ref{OMW}) have the obvious properties:
\begin{equation}
k\wedge{}^*\Theta = 0,\qquad k\wedge{}^*\Omega^\alpha = 0,\qquad 
k_\alpha\Omega^\alpha = 0.\label{kTOM}
\end{equation}
In accordance with (\ref{Wa0}), we have explicitly: $\Omega^a = 0$ ($a = 0,1$) and 
\begin{equation}
\Theta = \vartheta^A\left({\frac 12}\partial_AU - \delta_{AB}W^B\right),\qquad
\Omega^A = \vartheta^B\partial_B W^A.\label{THOMa}
\end{equation}
Applying the transversal differential to (\ref{THW}), and making use of (\ref{OMW}), we find
\begin{equation}
\underline{d}\Theta = \Omega_\alpha\wedge\vartheta^\alpha.\label{dTHW}
\end{equation}
In essence, this is equivalent to the Bianchi identity $DT^\alpha = R_\beta{}^\alpha\wedge\vartheta^\beta$ which is immediately checked by applying the covariant differential $D$ to (\ref{torW}) and using (\ref{curW}). 

Let us discuss the properties of the torsion and the curvature for the wave ansatz (\ref{cof0})-(\ref{cof23}) and (\ref{conW}). To begin with, it is worthwhile to notice that the 2-forms of the gravitational Ponicar\'e gauge field strengths (\ref{torW}) and (\ref{curW}) have the same structure as the electromagnetic field strength (\ref{Fij}) of a plane wave. Indeed, we have
\begin{equation}
T^\alpha = k\wedge a^\alpha,\qquad R^{\alpha\beta} = k\wedge a^{\alpha\beta},\label{poinW}
\end{equation}
where $a^\alpha = -\, k^\alpha\Theta$ and $a^{\alpha\beta} = -\,2k^{[\alpha}\Omega^{\beta]}$ play the role of the gravitational (translational and rotational, respectively) ``polarization'' 1-forms, in complete analogy to the polarization 1-form $a$ in (\ref{Fij}). Similarly to (\ref{ai}), the polarization 1-forms satisfy the orthogonality relations
\begin{eqnarray}
k\wedge{}^\ast a^\alpha = 0,\qquad k\wedge{}^\ast a^{\alpha\beta} = 0,\label{aka1}\\
k_\alpha a^\alpha = 0,\qquad k_\alpha a^{\alpha\beta} = 0.\label{aka2}
\end{eqnarray}
Clearly, the gravitational field strengths of a wave have the properties
\begin{eqnarray}
k\wedge{}^\ast\!T^\alpha = 0,\qquad k\wedge{}^\ast\!R^{\alpha\beta} = 0,\label{kTW}\\
k\wedge T^\alpha = 0,\qquad k\wedge R^{\alpha\beta} = 0,\label{kRW}\\
T^\alpha\wedge{}^\ast\!T^\beta = 0,\qquad  
R^{\alpha\beta}\wedge{}^\ast\!R^{\rho\sigma} = 0,\label{kTRW}
\end{eqnarray}
in complete analogy to the electromagnetic plane wave (\ref{kF1}).

In addition, however, the gravitational field strengths satisfy 
\begin{equation}
k_\alpha\,T^\alpha = 0,\qquad k_\alpha\,R^{\alpha\beta} = 0,\label{kTRW2}
\end{equation}
in view of (\ref{kTOM}).

\subsection{Irreducible decomposition of gravitational field strengths}

It is straightforward to find the irreducible parts of the torsion and the curvature. One can prove that the second (trace) and third (axial trace) irreducible parts of the torsion are trivial, ${}^{(2)}\!T^\alpha = 0$ and ${}^{(3)}\!T^\alpha = 0$, whereas 
\begin{equation}
{}^{(1)}\!T^\alpha = T^\alpha = -\,k\wedge k^\alpha\,\Theta.\label{torW1}
\end{equation}
At the same time, the curvature pieces ${}^{(3)}\!R^{\alpha\beta} = {}^{(5)}\!R^{\alpha\beta} =
{}^{(6)}\!R^{\alpha\beta} = 0$, and for $I = 1,2,4$:
\begin{equation}
{}^{(I)}\!R^{\alpha\beta} = 2\,k\wedge {}^{(I)}\!\Omega^{[\alpha}k^{\beta]},\label{curW124}
\end{equation}
where ${}^{(1)}\!\Omega^\alpha + {}^{(2)}\!\Omega^\alpha + {}^{(4)}\!\Omega^\alpha = \Omega^\alpha$, and
explicitly we find
\begin{eqnarray}
{}^{(1)}\!\Omega^\alpha &=& {\frac 12}\left(\Omega^\alpha - \vartheta^\alpha e_\beta\rfloor\Omega^\beta
+ \vartheta^\beta e^\alpha\rfloor\Omega_\beta\right),\label{OM1}\\
{}^{(2)}\!\Omega^\alpha &=& {\frac 12}\left(\Omega^\alpha - \vartheta^\beta e^\alpha\rfloor
\Omega_\beta\right),\label{OM2}\\
{}^{(4)}\!\Omega^\alpha &=& {\frac 12}\,\vartheta^\alpha e_\beta\rfloor\Omega^\beta.\label{OM4}
\end{eqnarray}
The transversal components of these objects are constructed in terms of the irreducible pieces of the $2\times 2$ matrix $\partial_BW^A$: symmetric traceless part, skew-symmetric part and the trace, respectively. Using (\ref{THOMa}), we derive ${}^{(I)}\!\Omega^A = {}^{(I)}\!\Omega^A{}_B\,\vartheta^B$, with
\begin{eqnarray}
{}^{(1)}\!\Omega^A{}_B &=& {\frac 12}\left(\partial_BW^A + \partial^AW_B - \delta^A_B
\,\partial_CW^C\right),\label{OMAB1}\\
{}^{(2)}\!\Omega^A{}_B &=& {\frac 12}\left(\partial_BW^A - \partial^AW_B\right),\label{OMAB2}\\
{}^{(4)}\!\Omega^A{}_B &=& {\frac 12}\,\delta^A_B\,\partial_CW^C.\label{OMAB4}
\end{eqnarray}

One can demonstrate the following properties of these 1-forms:
\begin{eqnarray}\label{vOM}
\vartheta_\alpha\wedge{}^{(1)}\!\Omega^\alpha = 0,\quad \vartheta_\alpha\wedge{}^{(2)}\!\Omega^\alpha 
= \vartheta_\alpha\wedge\Omega^\alpha,\\
\vartheta_\alpha\wedge{}^{(4)}\!\Omega^\alpha = 0,\quad
e_\alpha\rfloor{}^{(1)}\!\Omega^\alpha = -\,e_\alpha\rfloor\Omega^\alpha,\\ 
e_\alpha\rfloor{}^{(2)}\!\Omega^\alpha = 0,\quad e_\alpha\rfloor{}^{(4)}\!\Omega^\alpha = 
2e_\alpha\rfloor\Omega^\alpha,\label{eOM}\\
k_\alpha{}^{(1)}\!\Omega^\alpha = -\,{\frac 12}\,k\,e_\alpha\rfloor\Omega^\alpha,\quad
k_\alpha{}^{(2)}\!\Omega^\alpha = 0,\label{kOM}\\
k_\alpha{}^{(4)}\!\Omega^\alpha = {\frac 12}\,k\,e_\alpha\rfloor\Omega^\alpha,
\quad k\wedge{}^*{}^{(2)}\!\Omega^\alpha = 0,\\
k\wedge{}^*{}^{(1)}\!\Omega^\alpha = -\, k\wedge{}^*{}^{(4)}\!\Omega^\alpha = 
-\,{\frac 12}\,k^\alpha\,\vartheta_\beta\wedge{}^*\Omega^\beta.\label{kdOM}
\end{eqnarray}

\section{Field equations}\label{FE}

We now finally turn to the analysis of the quadratic Poincar\'e gauge model with the general Lagrangian (\ref{LRT}). We however assume the vanishing cosmological constant. Gravitational wave solutions for the case of a nontrivial cosmological constant require a special investigation. 

Substituting the torsion (\ref{torW1}) and the curvature (\ref{curW124}), into 
(\ref{HlaQT}) and (\ref{hR}), we find 
\begin{equation}
h^\alpha = -\,k^\alpha Z,\qquad h^{\alpha\beta} = -\,2k^{[\alpha}Z^{\beta]},\label{hZ}
\end{equation}
where we introduced the 2-forms 
\begin{eqnarray}
Z &=& a_1\,{}^*\!\left(k\wedge\Theta\right) + \overline{a}_1\,k\wedge\Theta,\label{ZT}\\
Z^\alpha &=& \sum\limits_{I=1,2,4} \left[b_I\,{}^*\!\left(k\wedge{}^{(I)}\!\Omega^\alpha\right) 
+ \overline{b}_I\,k\wedge{}^{(I)}\!\Omega^\alpha\right].\label{ZR}
\end{eqnarray}
Making use of (\ref{kTOM}) and (\ref{vOM})-(\ref{kdOM}) we can show that
\begin{eqnarray}
k\wedge h^\alpha = 0,\qquad k\wedge {}^*h^\alpha = 0,\qquad k_\alpha h^\alpha = 0,\label{kha}\\
k\wedge h^{\alpha\beta} = 0,\qquad k\wedge {}^*h^{\alpha\beta} = 0,
\qquad k_\alpha h^{\alpha\beta} = 0.\label{khab}
\end{eqnarray}
As a result, substituting (\ref{hZ}) into (\ref{qa}) and (\ref{qaR}), we find
\begin{equation}
q^{(T)}_\alpha = 0,\qquad q^{(R)}_\alpha = 0.\label{qqW}
\end{equation}
With an account of the properties (\ref{kha}) and (\ref{khab}), one can check that
\begin{equation}
Dh^\alpha = dh^\alpha,\qquad Dh^{\alpha\beta} = dh^{\alpha\beta}.\label{dhW}
\end{equation}

The transversal nature of $\Theta$ and $\Omega^A$ leads to a further simplification. In particular, using (\ref{kphi}), we recast (\ref{ZT}) and (\ref{ZR}) into
\begin{equation}
Z = k\wedge\Xi,\qquad Z^A = k\wedge\Xi^A,\label{ZZA}
\end{equation}
where we have introduced the 1-forms
\begin{eqnarray}
\Xi &=& a_1\,{}^{\underline{*}}\Theta + \overline{a}_1\,\Theta,\label{xiW}\\
\Xi^A &=& \sum\limits_{I=1,2,4} \left[b_I\,{}^{\underline{*}}{}^{(I)}\!\Omega^A 
+ \overline{b}_I\,{}^{(I)}\!\Omega^A\right].\label{xiAW}
\end{eqnarray}
Although for $a =0,1$ the components $Z^a$ are nontrivial, they are irrelevant. Indeed, one can verify that ($a,b = 0,1$ and $A,B = 2,3$)
\begin{equation}
h^{ab} = 0,\qquad h^{AB} = 0,\qquad h^{aB} = -\,k\wedge k^a\,\Xi^B,\label{habW}
\end{equation}
and consequently $Z^a$ does not enter the field equations. 

\begin{widetext}

\subsection{Wave equations}

After all these preparations, we are in a position to write down the gravitational field equations for the quadratic Poincar\'e gauge model. Substituting the gravitational wave ansatz into (\ref{ERT1}) and (\ref{ERT2}), we derive in vacuum (without matter sources, ${\mathfrak T}_\alpha = 0$ and ${\mathfrak S}^\alpha{}_\beta = 0$)
\begin{eqnarray}
-\,k_\alpha\,k\wedge\left\{a_0\,\vartheta_\beta\wedge{}^{\underline{*}}\Omega^\beta - {\frac 1\xi}
\,\vartheta_\beta\wedge\Omega^\beta + \underline{d}\,\Xi\right\} = 0,\label{EPW1}\\
k_a\,k\wedge\left\{(a_0 + a_1)\,\vartheta_B\wedge{}^{\underline{*}}\Theta + \Bigl({\frac 1\xi} + \overline{a}_1
\Bigr)\,\vartheta_B\wedge\Theta - 2\ell_\rho^2\,\underline{d}\,\Xi_B\right\} = 0.\label{EPW2}
\end{eqnarray}
Note that in the second field equation (\ref{ERT2}), the $[ab]$ and $[AB]$ components are satisfied identically, and only the mixed $[aB]$ components give rise to (\ref{EPW2}). This is completely consistent with (\ref{habW}). 

Making use of (\ref{xiW}) and (\ref{dTHW}), with an account of the traversal nature of $\Omega^\alpha$, we bring (\ref{EPW1}) and (\ref{EPW2}) to the form
\begin{eqnarray}
a_0\,\vartheta_A\wedge{}^{\underline{*}}\Omega^A - \Bigl({\frac 1\xi} + \overline{a}_1\Bigr)
\,\vartheta_A\wedge\Omega^A + a_1\,\underline{d}\,{}^{\underline{*}}\Theta = 0,\label{PW1}\\
(a_0 + a_1)\,\vartheta^A\wedge{}^{\underline{*}}\Theta + \Bigl({\frac 1\xi} + \overline{a}_1
\Bigr)\,\vartheta^A\wedge\Theta - 2\ell_\rho^2\,\underline{d}\,\Xi^A = 0.\label{PW2}
\end{eqnarray}
Both equations are 2-forms on the 2-dimensional transversal space spanned by $x^A = (x^2, x^3)$, and thus (\ref{PW1}) and (\ref{PW2}) is a system of three partial differential equations for the three variables $U = U(\sigma, x^B)$ and $W^A = W^A(\sigma, x^B)$. Substituting (\ref{THOMa}), we recast (\ref{PW1}) and (\ref{PW2}) into the final tensorial form
\begin{eqnarray}
(a_0 + a_1)\,\partial_AW^A - \Bigl({\frac 1\xi} + \overline{a}_1\Bigr)\,\eta^{AB}
\partial_A\underline{W}{}_B - {\frac {a_1}{2}}\,\underline{\Delta}\,U &=& 0,\label{W1}\\
(a_0 + a_1)\,\Bigl({\frac 12}\partial_AU - \underline{W}{}_A\Bigr)  - \Bigl({\frac 1\xi} + 
\overline{a}_1\Bigr)\,\eta_{AB}\Bigl({\frac 12}\underline{\partial}{}^BU - W^B\Bigr) &&\nonumber\\
+\, \ell_\rho^2\,\Big\{- (b_1 + b_2)\,\underline{\Delta}\,\underline{W}{}_A + (b_2 - b_4)
\,\partial_A\partial_BW^B && \nonumber\\
+\, (\overline{b}_1 - \overline{b}_2)\left(\partial_A\eta^{BC}\partial_B\underline{W}{}_C + 
\eta_{AB}\underline{\partial}{}^B\partial_CW^C\right)\Big\} &=& 0.\label{W2}
\end{eqnarray}
Here $\underline{\Delta} = \delta^{AB}\partial_A\partial_B$ is the 2-dimensional Laplacian on the transversal space, and we denote $\underline{W}{}_A = \delta_{AB}W^B$ and $\underline{\partial}{}^A = \delta^{AB}\partial_B$. 
\end{widetext}

The system (\ref{W1})-(\ref{W2}) always admits a nontrivial solution for the arbitrary quadratic Poincar\'e gauge model with any choice of coupling constants. There are some interesting special cases.

\subsection{Torsionless gravitational waves}

The torsion (\ref{torW}) vanishes when $\Theta = 0$ which is realized for 
\begin{equation}
W^A = {\frac 12}\delta^{AB}\partial_BU.\label{notorW}
\end{equation}
Substituting this into (\ref{W1}), we find 
\begin{equation}
{\frac {a_0}{2}}\,\underline{\Delta}\,U = 0,\label{notW1}
\end{equation}
whereas (\ref{W2}) reduces to 
\begin{equation}
\ell_\rho^2\,\underline{\partial}{}^B\left\{- (b_1 + b_4)\delta_{AB}\,\underline{\Delta}\,U
+ (\overline{b}_1 - \overline{b}_2)\,\eta_{AB}\,\underline{\Delta}\,U\right\} = 0.\label{notW2}
\end{equation}
Accordingly, we conclude that the well-known torsionless wave solution of GR with the function $U$ satisfying the Laplace equation is an exact solution of the generic quadratic Poincar\'e gauge gravity model. This is consistent with our earlier results on the torsion-free solutions in Poincar\'e gauge theory \cite{selected,Obukhov:1989}.

Moreover, the torsionless wave (\ref{notorW})-(\ref{notW1}) represents a general solution for the purely torsion quadratic class of  Poincar\'e models, since this is the only configuration admitted by the system (\ref{W1})-(\ref{W2}) for $b_I = \overline{b}_I = 0$.

\subsection{Teleparallel gravitational waves}

The curvature (\ref{curW}) vanishes for $\Omega^\alpha = 0$ which is realized when $W^\alpha = W^\alpha(\sigma)$ is independent of the transversal coordinates. Such a solution only exists in a class of Poincar\'e gauge models restricted by the conditions on the coupling constants
\begin{equation}
a_0 + a_1 = 0,\qquad {\frac 1\xi} + \overline{a}_1 = 0.\label{teleW}
\end{equation}
The system (\ref{W1})-(\ref{W2}) then reduces to
\begin{equation}
{\frac {a_1}{2}}\,\underline{\Delta}\,U = 0,\label{teleW1}
\end{equation}
Accordingly, the metric structure turns out to be the same for the torsionless and teleparallel gravitational wave solutions.

\subsection{Torsion gravitational waves}

The torsion-free ansatz (\ref{notorW}) can be generalized to 
\begin{equation}
W^A = {\frac 12}\left(\delta^{AB}\partial_BV + \eta^{AB}\partial_B\overline{V}\right),\label{pW}
\end{equation}
with $V \neq U$. Substituting this into (\ref{W1}) and (\ref{W2}), we derive
\begin{eqnarray}
(a_0 + a_1)\,\underline{\Delta}\,V + \Bigl({\frac 1\xi} + \overline{a}_1\Bigr)\,
\underline{\Delta}\,\overline{V} - a_1\,\underline{\Delta}\,U &=& 0,\label{pW1}\\
\partial_A\Big\{(a_0 + a_1)(U - V) -  \Bigl({\frac 1\xi} + \overline{a}_1\Bigr)\,
\overline{V}\nonumber\\
-\,\ell_\rho^2\left[(b_1 + b_4)\,\underline{\Delta}\,V + (\overline{b}_1 - \overline{b}_2)
\,\underline{\Delta}\,\overline{V}\,\right]\Big\} && \nonumber\\
-\,\eta_{AB}\underline{\partial}{}^B\Big\{(a_0 + a_1)\,\overline{V} + \Bigl({\frac 1\xi} 
+ \overline{a}_1\Bigr)\,(U - V)\nonumber\\
+\,\ell_\rho^2\left[(b_1 + b_2)\,\underline{\Delta}\,\overline{V} - (\overline{b}_1 
- \overline{b}_2)\,\underline{\Delta}\,V\right]\Big\} &=& 0.\label{pW2}
\end{eqnarray}

As a result, we obtain the system of the three linear second order differential equations for the three functions $U, V, \overline{V}$:
\begin{eqnarray}
(a_0 + a_1)\,\underline{\Delta}\,V + \Bigl({\frac 1\xi} + \overline{a}_1\Bigr)\,
\underline{\Delta}\,\overline{V} - a_1\,\underline{\Delta}\,U &=& 0,\label{LW1}\\
\ell_\rho^2\left[(b_1 + b_4)\,\underline{\Delta}\,V + (\overline{b}_1 - \overline{b}_2)
\,\underline{\Delta}\,\overline{V}\,\right] && \nonumber\\
-\,(a_0 + a_1)(U - V) + \Bigl({\frac 1\xi} + \overline{a}_1\Bigr)\,\overline{V} &=& 0,\label{LW2}\\
\ell_\rho^2\left[(b_1 + b_2)\,\underline{\Delta}\,\overline{V} - (\overline{b}_1 
- \overline{b}_2)\,\underline{\Delta}\,V\right]\nonumber\\
+\,(a_0 + a_1)\,\overline{V} + \Bigl({\frac 1\xi} + \overline{a}_1\Bigr)\,(U - V) &=& 0.\label{LW3}
\end{eqnarray}
The structure of (\ref{LW1})-(\ref{LW3}) suggests that it is more convenient to work with the set of variables $(U - V), V, \overline{V}$. This is a coupled system of Helmholtz or screened Laplace equations and it can be straightforwardly solved. Since this is a linear system, solution can be sought in the form
\begin{equation}
\left(\begin{array}{c}U - V \\ V \\ \overline{V}\end{array}\right) = e^{iq_Ax^A} 
\left(\begin{array}{c}U_0 - V_0 \\ V_0 \\ \overline{V}_0\end{array}\right),\label{eqx}
\end{equation}
with constant $q_A$. Of course, in the end one should take the real part of (\ref{eqx}). Depending on whether $q_A$'s are real or imaginary, one finds oscillating or exponentially decaying/growing configurations. Substituting the ansatz (\ref{eqx}), into (\ref{LW1})-(\ref{LW3}), we derive an algebraic system for the constant parameters:
\begin{widetext}\begin{equation}\label{q2eq}
\left(\begin{array}{ccc}-a_1 & a_0 & {\frac 1\xi} + \overline{a}_1 \\
-(a_0 + a_1) & - Q^2(b_1 + b_4) & {\frac 1\xi} + \overline{a}_1 
- Q^2(\overline{b}_1 - \overline{b}_2) \\ {\frac 1\xi} + \overline{a}_1 & 
Q^2(\overline{b}_1 - \overline{b}_2) & a_0 + a_1 - Q^2(b_1 + b_2) 
\end{array}\right)\left(\begin{array}{c}U_0 - V_0 \\ V_0 \\ \overline{V}_0\end{array}\right) = 0,
\end{equation}
where we denoted $q^2 := \delta^{AB}q_Aq_B$ and introduced dimensionless  $Q^2 := q^2\ell_\rho^2$. 
A nontrivial solution exists when the determinant vanishes, which yields a quadratic equation for 
$Q^2$: 
\begin{equation}
{\mathcal A}Q^4 + {\mathcal B}Q^2 + {\mathcal C} = 0,\label{ABCQ}
\end{equation}
where the coefficients read explicitly
\begin{eqnarray}\label{AQ}
{\mathcal A} &=& -\,a_1\left[(b_1 + b_4)(b_1 + b_2) + (\overline{b}_1 - \overline{b}_2)^2\right],\\
{\mathcal B} &=& (a_0 + a_1)\left[a_1(b_1 + b_4) - a_0(b_1 + b_2)\right]\nonumber\\
&& + \Big({\frac 1\xi} + \overline{a}_1\Big)^2 (b_1 + b_4) - 2a_0\,\Big({\frac 1\xi} 
+ \overline{a}_1\Big)(\overline{b}_1 - \overline{b}_2),\label{BQ}\\
{\mathcal C} &=& a_0\left[(a_0 + a_1)^2 + \Big({\frac 1\xi} + \overline{a}_1\Big)^2\right].\label{CQ}
\end{eqnarray}
\end{widetext}
The resulting set of the gravitational wave solutions encompasses both oscillating and exponentially decaying/growing profiles, the form of which is fully determined by the structure of the gravitational Lagrangian (in technical terms, by the values of the coupling constants). 

In particular, for the class of the Yang-Mills type models with the purely curvature quadratic Lagrangians (with the trivial coupling constants $a_0 = 0$, ${\frac 1\xi} = 0$, $a_I =  0$, and $\overline{a}_I = 0$), 
the system (\ref{LW1})-(\ref{LW3}) yields 
\begin{equation}
\underline{\Delta}\,V = 0,\qquad \underline{\Delta}\,\overline{V} = 0,\label{DVV}
\end{equation}
whereas $U$ is an arbitrary function.

\section{Discussion and conclusion}\label{DC}

In this paper we have derived the plane-fronted gravitational waves for the class of the Poincar\'e gravity models with the most general Lagrangian (\ref{LRT}) which includes all possible linear and quadratic invariants of the torsion and the curvature. Both, the parity even and the parity odd terms are taken into account. The exact solutions are obtained by making use of the properties of the electromagnetic plane waves. We demonstrate the existence of the torsion-free wave configurations of Einstein's GR (which reduce to the gravitational plane waves in the Brinkmann-Rosen form), as well as gravitational waves with the propagating torsion. The latter are characterized, in general, by the three profile wave variables $U, V, \overline{V}$ which satisfy the coupled system of the Helmholtz or screened Laplace equations. The form of the wave profile depends on the values of the coupling constants entering the Lagrangian: in physical terms, on the particle spectrum of the corresponding Poincar\'e gravity model \cite{Diakonov,Baekler1,Baekler2,Karananas}.

The Lagrangian (\ref{LRT}) can be further generalized if we add the cosmological constant term. However, already in GR the construction of the gravitational waves for the case of a nontrivial cosmological constant represents a difficult problem \cite{ndim}, and one needs a special investigation of this case for Poincar\'e gravity theory. Recently, the existence of the generalized Vaidya and Siklos gravitational waves was demonstrated in three and four dimensions \cite{BC1,BC2,BC3,BC4}. These solutions belong to a class of waves different from those considered in the present paper. The plane waves in the Poincar\'e gravity theory with a nontrivial cosmological constant were reported in \cite{BC5}. 

The plane gravitational waves are of fundamental importance in general relativity; it is worthwhile to recall that any Riemannian spacetime has a plane wave structure in a certain limit \cite{pen2}. In the framework of the gauge approach to gravity, the spacetime geometry becomes more complicated and it is thus of considerable interest to explore the possibility of extension of the general-relativistic results to the case of the non-Riemannian geometry. The systematic analysis presented here demonstrates the existence of a direct generalization of the plane waves from general relativity to the  Poincar\'e gravity theory. The new exact solutions have all the basic properties of electromagnetic and general-relativistic gravitational plane waves. The results obtained are valid for the widest possible class of quadratic models which suggests their further adaptation and application in the modified gravity models with Lagrangians arbitrarily depending on the curvature and the torsion. 

\begin{acknowledgments}
I would like to thank Milutin Blagojevi\'c for the correspondence, kindly informing about his recent work. This work was partially supported by the Russian Foundation for Basic Research (Grant No. 16-02-00844-A).
\end{acknowledgments}
\bigskip

\appendix

\section{Irreducible decomposition of the torsion and curvature}\label{irreducible}

The torsion 2-form can be decomposed into the three 
irreducible pieces, $T^{\alpha}={}^{(1)}T^{\alpha} + {}^{(2)}T^{\alpha} + 
{}^{(3)}T^{\alpha}$, where
\begin{eqnarray}
{}^{(2)}T^{\alpha}&=& {\frac 13}\vartheta^{\alpha}\wedge (e_\nu\rfloor 
T^\nu),\label{iT2}\\
{}^{(3)}T^{\alpha}&=& {\frac 13}e^\alpha\rfloor(T^{\nu}\wedge
\vartheta_{\nu}),\label{iT3}\\
{}^{(1)}T^{\alpha}&=& T^{\alpha}-{}^{(2)}T^{\alpha} - {}^{(3)}T^{\alpha}.
\label{iT1}
\end{eqnarray}
The Riemann-Cartan curvature 2-form is decomposed $R^{\alpha\beta} = 
\sum_{I=1}^6\,{}^{(I)}\!R^{\alpha\beta}$ into the 6 irreducible parts 
\begin{eqnarray}
{}^{(2)}\!R^{\alpha\beta} &=& -\,{}^*(\vartheta^{[\alpha}\wedge\Psi^{\beta]}),
\label{curv2}\\
{}^{(3)}\!R^{\alpha\beta} &=& -\,{\frac 1{12}}\,{}^*(X\,\vartheta^\alpha\wedge
\vartheta^\beta),\label{curv3}\\
{}^{(4)}\!R^{\alpha\beta} &=& -\,\vartheta^{[\alpha}\wedge\Phi^{\beta]},
\label{curv4}\\
{}^{(5)}\!R^{\alpha\beta} &=& -\,{\frac 12}\vartheta^{[\alpha}\wedge e^{\beta]}
\rfloor(\vartheta^\gamma\wedge W_\gamma),\label{curv5}\\
{}^{(6)}\!R^{\alpha\beta} &=& -\,{\frac 1{12}}\,W\,\vartheta^\alpha\wedge
\vartheta^\beta,\label{curv6}\\
{}^{(1)}\!R^{\alpha\beta} &=& R^{\alpha\beta} -  
\sum\limits_{I=2}^6\,{}^{(I)}R^{\alpha\beta},\label{curv1}
\end{eqnarray}
where 
\begin{eqnarray}
W^\alpha := e_\beta\rfloor R^{\alpha\beta},\quad W := e_\alpha\rfloor W^\alpha,
\label{WX1}\\
X^\alpha := {}^*(R^{\beta\alpha}\wedge\vartheta_\beta),\quad X := 
e_\alpha\rfloor X^\alpha,\label{WX2}
\end{eqnarray}
and 
\begin{eqnarray}
\Psi_\alpha &:=& X_\alpha - {\frac 14}\,\vartheta_\alpha\,X - {\frac 12}
\,e_\alpha\rfloor (\vartheta^\beta\wedge X_\beta),\label{Psia}\\
\Phi_\alpha &:=& W_\alpha - {\frac 14}\,\vartheta_\alpha\,W - {\frac 12}
\,e_\alpha\rfloor (\vartheta^\beta\wedge W_\beta)\label{Phia}.
\end{eqnarray}

\end{document}